\documentclass[12pt]{article}
\title{
Baryon magnetic moments in the QCD string approach\\
 }
\author{B.O. Kerbikov, Yu.A.Simonov\\ State Research
Center\\Institute of Theoretical and Experimental Physics, \\ Moscow,
Russia}
 \date{}
  \newcommand{\be}{\begin{equation}}
\newcommand{\ee}{\end{equation}}
 
\def\fun#1#2{\lower3.6pt\vbox{\baselineskip0pt\lineskip.9pt
\ialign{$\mathsurround=0pt#1\hfil
##\hfil$\crcr#2\crcr\sim\crcr}}}
\newcommand{\vesig}{\mbox{\boldmath${\rm \sigma}$}}

 \newcommand{\lan}{\langle}
\newcommand{\ran}{\rangle}

\begin{document}
\maketitle

\begin{abstract}
Magnetic moments of baryons composed of light and strange quarks
are computed for the first time through the only  parameter of the
model-- string tension $\sigma$.  For example, $\mu_p=
m_p/c\sqrt{\sigma},$
$\mu_{\Omega^-}=-\mu_p(1+\frac34
\frac{m^2_s}{c^2\sigma}-\frac{15}{32}
\frac{m^4_s}{c^4\sigma^2})^{-1}$, where $m_p$ is the  proton
mass,  $m_s$ is the strange quark current mass, $c=0.957$  -- a
constant which is caluculated in the paper.  Resulting theoretical
values  differ from the experimental  ones typically by about
 10\%.  \end{abstract}

\section{Introduction} Recently in a  stimulating  analysis of the
hyperon static properties \cite{0.1} H.J.Lip\-kin displayed remarkably
successful relations connecting strange and  nonstrange baryon
magnetic moments and the corresponding quark masses $\nu_u$ and
$\nu_s$ (in order to avoid similar notations for masses and magnetic
moments we denote current quark masses by $m_k$ and dynamical,  or
constituent, quark masses by $\nu_k$).  To understand why these
relations so well agree with the experiment and to  get insight into
the problems encountered in semileptonic decays of baryons \cite{0.1}
one needs a dynamical approach which would enable one  to express
constituent masses of quarks and baryon magnetic moments in terms of
a single QCD scale parameter.  It is a purpose of the present letter
to express the magnetic moments of  baryons and constituent
masses of the corresponding quarks in terms of only one parameter --
the string tension, and to demonstrate that our results are in  line
with the relations of Lipkin.

 Theoretical investigation of baryon magnetic moments (BMM) has a
long history \cite{1}.
 In the constituent quark model (CQM) BMM are expressed through the
 values of constituent quark masses, which are input parameters
 \cite{2} (see also \cite{3} for discussion). Among other approaches
 to the problem mention should be made of different versions of the
 bag model \cite{4}, lattice calculations \cite{5} and the QCD sum
 rules \cite{6}.  In the latter the BMM are connected to the values
 of chiral and gluonic condensates and to the  quartic quark
 correlator. Although a lot of efforts has been undertaken along
 different lines, the theoretical predictions still differ from the
 experimental values (by 10-15\% in the worst case \cite{7}).

 In all models however theoretical predictions are somewhat biased by
 the introduction of supplementary parameters in addition to the only
 one pertinent to QCD -- the overall scale of the theory, which
 should be specified to make the QCD complete. In the final, ideal
 case this  role is played by $\Lambda_{QCD}$; in our treatment, as
 well as in lattice QCD calculations, we take as this universal
 parameters the QCD string tension $\sigma$, fixed in nature by the
 meson  and baryon  Regge slopes.

 The purpose of our letter is to calculate BMM through this single
parameter, $\sigma$, in the simplest possible approximation within
the nonperturbative QCD approach, developed in \cite{8}-\cite{13}.

 \section{Relativistic $3q$  Green's function and effective
 Hamiltonian}

 The starting point of the approach is the
 Feynman--Schwinger (world-line) representation of the
 $3q$ Green's function \cite{8}, where the role of "time"
 parameter along the path $z^{(i)}_\mu(s_i)$ of $i$-th
 quark is the Fock--Schwinger proper time $s_i$,
 $i=1,2,3$. One has \cite{8},\cite{10}
 \be
 G^{(3q)}(X,Y)=\int \prod^3_{i=1} ds_i Dz^{(i)}_\mu
 e^{-K}\lan W_3(X,Y)\ran
 \label{1}
 \ee
 where $X;Y =x^{(1)}, x^{(2)}, x^{(3)}; y^{(1)},
 y^{(2)}, y^{(3)}$,
 \be
 K=\sum^3_{i=1} (m^2_is_i+
 \frac{1}{4}\int^{s_i}_0\left
 (\frac{dz^{(i)}_\mu}{d\tau_i}\right)^2d\tau_i).
 \label{2}
 \ee
 Here $m_i$ is the current quark mass
 and the three-lobes Wilson loop is a product of three
 parallel transporters
 \be
 \lan W_3(X,Y)\ran =\lan \prod^3_{i=1} \Phi^{(i)}_{a_ib_i} (x^{(i)},
 y^{(i)})\ran e_{a_1a_2a_3} e_{b_1b_2b_3}.
 \label{3}
 \ee
  The standard approximation in the QCD string approach is the
  minimal area law for (\ref{3}), which will be used in what follows
  \be
  \lan W_3\ran=exp (- \sigma\sum^3_{i=1} S_i)
  \label{4}
  \ee
  where $S_i$ is the minimal area of one loop.

  The next step is basic for our approach, and it  allows finally to
  calculate the quark constituent masses $\nu_i$ in terms of the
  quark current masses $m_i$, defined at the  scale of $1 GeV$.

In this step one connects proper and real times (in the baryon c.m.
system)
\be
ds_i=\frac{dt}{2\nu_i(t)}
\label{5}
\ee
where $t=z_4^{(i)}(s_i), 0\leq t\leq T,$ is a common c.m. time on the
hypersurface $t=const.$   The new entity, $\nu_i(t)$ as will be seen,
plays the role of the quark constituent mass and will be calculated
through $\sigma$(and $\alpha_s$ when perturbative exchanges are taken
into account).

Considering the exponent in (\ref{3}) as an action one can define
the Hamiltonian and go over to the representation  \cite{9},\cite{12}
\be
G^{(3q)}=\int\prod^3_{i=1} D\nu_i(t)D^3z^{(i)} (t) e^{-A}
\label{6}
\ee
with
\be
A=\sum^3_{i=1} \int^T_0 dt \left(\frac{m^2_i}{2\nu_i}+ \frac{\nu_i}{2}
+\frac{ (\dot{\bf z}^{(i)}(t))^2}{2\nu_i}\right)+\sigma S_i
\label{7}
\ee

The final step in the approach \cite{10}, \cite{11}
\cite{12} is the derivation of the
c.m.  Hamiltonian containing $\nu_i$ as parameters to be found from
the condition of the Hamiltonian minimum.
It has the following form
\be
H=\sum^3_{k=1}\left (\frac{m^2_k}{2\nu_k}+ \frac{\nu_k}{2}\right
)+ \frac{1}{2m}
\left (-\frac{\partial^2}{\partial {\bf \xi}^2}-
\frac{\partial^2}{\partial {\bf \eta}^2}\right)+
\sigma\sum^3_{k=1}|{\bf r}^{(k)}|.
\label{8}
\ee

Here ${\bf \xi}, {\bf \eta}$ are Jacobi coordinates defined as in
\cite{11}, and ${\bf r}^{(k)}$ is the distance from the $k$-th quark
to the string-junction position which  we take below for simplicity
coinciding with the c.m.point.
In addition (\ref{8}) contains an arbitrary mass parameter $m$
  introduced to ensure correct dimensions, this parameter drops out
 from final expressions. Leaving technical details to the Appendixes,
 we now treat the Hamiltonian (\ref{8}) using the hyperspherical
 formalism \cite{14}.

\section{
Evaluation of quarks constituent masses}

 Considering three quarks
with equal masses and introducing the hyperradius $\rho^2={\bf
\xi}^2+{\bf \eta}^2$, one has in the  approximation  of the lowest
hyperspherical harmonic (which is known \cite{15} to yield accuracy
of the eigenvalue $E_n$ around one percent) \be
\frac{d^2\chi(\rho)}{d\rho^2}+2\nu\{E_n-W(\rho)\}\chi(\rho)=0,
\label{9}
\ee
\be
W(\rho)=b
\rho+
\frac{d}{2\nu\rho^2}, b=\sigma\sqrt{\frac23}\frac{32}{5\pi}, d= 15/4.
\label{10a}
\ee

The baryon mass $M_n(\nu)$ is equal to (for equal quark masses)
\be
M_n(\nu)=\frac{3m^2}{2\nu}
+\frac32\nu + E_n(\nu).
\label{10}
\ee
The crucial  point is now   the calculation of  $\nu$, which
is to be found from the minimum of $M_n(\nu)$, as it is prescribed
in the QCD string approach \cite{8}-\cite{13}.

At this point it is important to stress that we have changed from
$\nu(t)$ depending on $t$ on the trajectory in the path integral
(\ref{6}) to  the operator $\nu$ to be found from momenta and
coordinates in (\ref{8}) as in
\cite{12} and finally to the constant
$\nu$ to be found from the minimum of the mass $M$, as it was
suggested in \cite{9},\cite{10}, \cite{13}. The accuracy of this
replacement was tested recently in \cite{16} to be around 5\% or
better for lowest levels.

The equations defining the stationary points of $M_n$ as function of
$\nu$ for equal masses is
\be
\frac{\partial M_n}{\partial
\nu}\left\vert_{\nu=\nu_{(0)}}=0\right.~~
\label{11}
\ee

The generalization for baryon made of three quarks with different
masses is straightforward.  The perturbative gluon exchanges and
spin--dependent terms can be selfconsistently included in the
above picture \cite{13}.  Including the Coulomb term and passing to
 dimensionless quantities $x, \varepsilon_n$ and $\lambda$ defined
 as \be
 x=(2\nu b)^{1/3} \rho, ~~ \varepsilon_n =\frac{2\nu E_n}{(2\nu
 b)^{2/3}},~~ \lambda=\alpha_s\frac83
 \left(\frac{10\sqrt{3}\nu^2}{\pi^2\sigma}\right)^{1/3},
 \label{13a}
 \ee
 where $\alpha_s$ is the strong coupling constant, one arrives at the
 following reduced equation
 \be
 \left\{-\frac{d^2}{dx^2}+x+\frac{d}{x^2}-\frac{\lambda}{
 x}-\varepsilon_n(\lambda)\right\} \chi(x)=0.
 \label{14a}
 \ee
 It is now a simple task to find eigenvalues $\varepsilon_n(\lambda)$
 of (\ref{14a}) either numerically, or analytically (see below). Then
 (\ref{11}) would yield the following equation
  defining the quark dynamical mass $\nu$
   \be
\varepsilon_n(\lambda)\left(
 \frac{\sigma}{\nu^2}\right)^{2/3}
 \left \{ 1+ \frac{2\lambda}{\varepsilon_n(\lambda)}\left
 \vert
 \frac{d\varepsilon_n}{d\lambda}\right\vert\right\}+\frac{9}{16}
 \left(\frac{75 \pi^2}{2}\right)^{1/3} \left
 (\frac{m^2}{\nu^2}-1\right )=0.
  \label{15a} \ee

 It turns out that numerical solution of (\ref{14a}) may be
   reproduced analytically
  with the
  accuracy of (1-2)\%
    provided one
 replaces  the potential $W(x)=x+d/x^2-\lambda/x$ in
 (\ref{14a}) by oscillator potential near the stationary
 point $W'(x_0)=0$ -- see the Appendix A.
   Equation (\ref{15a})  applied to the nucleon $(m=0)$ yields
 the dynamical mass $\nu_u$ of the light quark, and applied to
 $\Omega^- (m=m_s) $ gives the strange quark mass $\nu_s$.
 Before  presenting these solutions we remind about spin-spin forces
 responsible e.g. for $N-\Delta$ splitting.
 Contrary to what might be naively expected the inclusion of
 spin-spin interaction considerably simplifies the problem due to
 remarkable cancellation of Coulomb and spin-spin  contributions into
 the dynamical (constituent) quark mass -- see the Appendix A.
 Therefore these terms should be kept only if one wishes to calculate
 the BMM with the accuracy much higher than 10\% in which case one
 should also take into account pion corrections, higher
 hyperspherical harmonics, etc. which is out of the scope of the
 present paper.

 Thus in order to determine the quark masses and eventually the
 baryon magnetic moments one needs only two parameters: the string
 tension $\sigma$ and the strange quark current mass $m_s$ ($u,d$
 current quark masses are set to zero).
 Present calculations were performed for
  \be
  \sigma=0.15 GeV^2,~~ m_s =0.245 GeV.
  \label{16a}
  \ee
  The string tension value (16) which is smaller than in the meson
  case is in line with baryon calculations by Capstick and Isgur
  \cite{S.C}. A similar smaller value of $\sigma$ is implied by
  recent lattice calculations by Bali \cite{G.B}.  Since the value of
   the above parameters are allowed to vary within certain limits
  \cite{13},\cite{G.B},\cite{23} one can in principle formulate the
  inverse problem, namely express the BMM in line with the present
  work and then fit their experimental values to determine the
  optimal choice of $\sigma$ and $m_s$.

  Consider first the case of  a nucleon made of three quarks with
 zero current masses and equal dynamical masses $\nu_u$. Keeping in
 mind cancellation of the Coulomb and spin-spin terms and thus
 setting in (\ref{14a}) and (\ref{15a}) $\lambda=0$ and making use of
 the oscillator approximation described in the Appendix A, one
 finds from (\ref{15a})
 \be
  \nu_u=2\sqrt{\frac{2\sigma}{\pi}}
 \left[\frac{2}{3\cdot 5^{1/3}}\left
 (1+\frac{2}{3\sqrt{5}}\right)\right]^{3/4} \equiv
 c\sqrt{\sigma} \simeq 0.957 \sqrt{\sigma} =0.37 GeV.
 \label{17a}
  \ee
  This result agrees with the exact solution of (\ref{14a}) at
 $\lambda=0$ with the accuracy better than 1\%. Similar  procedure
 applied to $\Omega^- $ baryon yields the strange quark dynamical
 mass $\nu_s$. From (\ref{15a}) and (\ref{17a}) one gets
 \be
 \nu_s\simeq c\sqrt{\sigma}\left( 1+\frac34
 \frac{m^2_s}{c^2\sigma} -\frac{15}{32}
 \frac{m^4_s}{c^4\sigma^2}\right) =0.46 GeV
 \label{18}
 \ee
 for $m_s=0.245 GeV$, and where the constant $c$ is defined in
 (\ref{17a}). Now we turn to baryon magnetic moments.

\section{Baryon magnetic moments}

 The form (\ref{1}) for
$G^{(3q)}$ does not take into account spins of quarks. When those are
inserted, a new additive term appears in the exponent of (\ref{6}),
proportional to the external magnetic field ${\bf B}$, namely $A$
acquires the following term \cite{17}, \cite{18}
 \be
 \delta
 A=\sum^3_{k=1}\int^{s_k}_0 d\tau_k e_k\vesig^{(k)}{\bf B}=
 \sum^3_{k=1}\int^T_0\frac{e_k{\vesig}^{(k)}{\bf
 B}}{2\nu_k} dt,
 \label{13}
 \ee
 where $e_k$ is the electric charge of the quark,
 ${\vesig}^{(k)}$ is the corresponding spin operator, and the
 definition (\ref{5}) of the constituent mass was used.

 Introducing the $z$-component of the magnetic moment operator
 \be
 \mu_z
= \sum^3_{k=1}\frac{e_k{\sigma}^{(k)}_z}{2\nu_k},
 \label{0.14}
 \ee
 one can write the BMM as matrix elements
 \be
 {\mu}_B\equiv \lan \Psi_B|
 {\mu}_z|\Psi_B\ran
 \label{14}
 \ee
 where $\Psi_B$ is the eigenfunction of (\ref{8}), and $\nu_k$ is
 taken at the stationary point,
 given by (\ref{11}).

For the baryon wave function we shall take here the simplest
approximation, namely
\be
\Psi_B=
\Psi^{symm}(r) \psi^{symm}(\sigma, f) \psi^a(color),
\label{15}
\ee
where $\psi(\sigma, f)$ is the spin-flavour part of the wave
function. The form (\ref{15}) neglects the  nonsymmetric components in
the coordinate $\psi(r)$ and spin-flavour parts of wave function,
which appear in the higher approximation of the hyperspherical
formalism \cite{14},  and for lowest states contribute only few
percent to the normalization \cite{14}.  The  spin-flavour
   functions  for different baryons were known for a
   long time \cite{2}, and  are  briefly outlined in the Appendix B.
   Using these functions it is a simple task to calculate the matrix
   element (\ref{14}), e.g.  the proton and neutron magnetic moments
   are given by
\be
\mu_p=\frac{m_p}{\nu_u}=\frac12\sqrt{\frac{\pi}{2\sigma}}\left[
\frac{2}{3\cdot 5^{1/3}}\left(1 +
\frac{2}{3\sqrt{5}}\right)\right]^{-3/4}\simeq 2.54 \mu_N,~~
\mu_n=-\frac23\mu_p\simeq -1.69 \mu_N.
\label{23}
\ee
Magnetic  moments of other baryons as well as new relations between
them are obtained from (\ref{15a}), (\ref{17a}) and (\ref{18}).

In particular one has
\be
\mu_{\Omega^-}\simeq -\mu_p(1+\frac34
\frac{m^2_s}{c^2\sigma}-\frac{15}{32}
\frac{m_s^4}{c^4\sigma^2})^{-1}=-2.04\mu_N,
\label{24}
\ee
\be
\mu_{\Sigma^+}(4c^2\sigma+ m^2_s)=-2\mu_{\Xi^0}(3c^2\sigma+2m^2_s),
\label{25}
\ee
where terms of the order $m^4_s/c^4\sigma^2$ were omitted in deriving
the last relation.

 Results on the BMM  are summarized in Table 1. As
these results differ  from the experimental values
typically  by only about 10\%
and are subjected to plentiful corrections (meson exchanges, higher
harmonics, etc.), one may conclude that the  outlined QCD approach is
successful even in its simplest form.

 It is important to realize that the QCD string model used above  is
 a fully relativistic string model for light current masses, and  the
 "nonrelativistic" appearence of the Hamiltonian (\ref{1}) is a
 consequence of the rigorous eibein formalism \cite{19} which was
 introduced in the most general form in \cite{20}.

 The approach enables to investigate other electromagnetic properties
      of baryons: transition magnetic  moments, polarizabilities
      \cite{25}, etc.

      The authors are grateful to Yu.S.Kalashnikova
      for numerous enlighting discussions and suggestions  and to
      A.M.Ba\-da\-lian and N.O.Agasian for useful remarks.

      The financial support of the grants RFFI 97-02-16406  and
      RFFI 96-15-96740
      are
      gratefully acknowledged, Yu.S. was partially supported by the
      RFFI grant 97-0217491.

        \newpage

{\bf Table 1.} Magnetic moments of baryons (in nuclear magnetons)
  computed using Eqs.(\ref{11}),(\ref{0.14}),(\ref{14}) in
  comparison with experimental data from PDG \cite{24} \begin{center}

\begin{tabular}{|l|l|l|l|l|l|l|l|l|l|}\hline
Baryon&$p$&$n$&
$\Lambda$&$\Sigma^-$&$\Sigma^0$&$\Sigma^+$&
$\Xi^-$&$\Xi^0$&$\Omega^-$\\\hline
Present work&2.54 &-1.69 &-0.69& -0.90 &0.80
&2.48&-0.63&-1.49&-2.04\\\hline
Experiment&2.79 &-1.91&-0.61&-1.16&&2.46&-0.65&-1.25&-2.02\\ \hline
 \end{tabular}
  \end{center}

      \newpage

\setcounter{equation}{0} \def\theequation{A.\arabic{equation}}

     \begin{center}
     {\bf Appendix A}\\

     {\bf Eigenvalue equation for baryons
     }\\
     \end{center}

     The eigenvalue equation for baryons in hyperspherical basis
     \cite{14} in its standard form is given by (\ref{9}) and reduced
     form with Coulomb term included by (\ref{14a}).
Though (\ref{14a}) can be easily solved numerically, it is
instructive to present the analytical solution resorting to the
oscillator approximation near the stationary point $W'(x_0)=0$. This
approximation is known to yield the accuracy of 1-2\% \cite{15}. It
will be demonstrated below  that the Coulomb term which tends to
increase the quark dynamical mass almost exactly  cancel with the
spin-spin interaction term. Therefore   one can consider (\ref{14})
at  $\lambda=0$. Then the ground state energy is equal to $W(x_0)$
plus the first quantum correction $\omega/2$, where $\omega^2=
2W^{\prime\prime}(x_0)$.

This yields
\be
\varepsilon (0) =\frac32(\frac{15}{2})^{1/3}(1+\frac{2}{3\sqrt{5}}).
\label{A.1}
\ee

Substitution of this result into (\ref{13a}) and
(\ref{10})-(\ref{11}) leads to the expression (\ref{17a}) for the
light quark current mass $\nu_u$.

Next we demonstrate the cancellation of the Coulomb and spin-spin
contributions into the quark mass. Again this conclusion results
directly from numerical calculations but it is always  preferable to
present transparent estimates. The derivative $dE_n/d\nu$ (see
(\ref{10}), (\ref{11}), (\ref{13a})) may be written as
\be
\frac{dE_n}{d\nu}=
\varepsilon(\lambda)
\frac{\partial}{\partial\nu}
\frac{(2\nu
b)^{2/3}}{2\nu}-
\frac{(2\nu
b)^{2/3}}{2\nu}
\frac{\partial\lambda}{\partial\nu}
\left\vert
\frac{d\varepsilon}{d\lambda}\right\vert,
\label{A.2}
\ee
where the (-) sign stems from the fact that
$d\varepsilon/d\lambda<0$.
Expanding $\varepsilon (\lambda)$ in Taylor series in $\lambda$ and
keeping only linear term (the small parameter is
$\lambda/\varepsilon(0) \simeq 1/4$) one gets
\be
\frac{dE_n}{d\nu}\simeq - \frac{(2\nu b)^{2/3}}{6\nu^2}
\varepsilon(0)\left\{1+\frac{\lambda}{\varepsilon(0)}\left\vert
\frac{d\varepsilon}{d\lambda}\right\vert\right\}.
\label{A.3}
\ee
The value of $\varepsilon(0)$ is given by (\ref{A.1}), the estimate
of $d\varepsilon/d\lambda$ in the small $\lambda$ regime is
strightforward, then recalling that according to (\ref{13a}) $\lambda
\propto (\nu^2/\sigma)^{1/3}$ and solving the simple equations one
obtains that due to Coulomb interaction $\nu_u$ increases by $\simeq
0.03 GeV$. This is confirmed by numerical solution of (\ref{14a}).

The spin-spin interaction in baryon results in the shift of $E_n$
equal to
\be
\delta E_n=\frac{16}{9} \alpha_s\sum_{i>j}\frac{{\bf s}_i{\bf
s}_j}{\nu_i\nu_j}\delta ({\bf r}_{ij}).
\label{A.4}
\ee
For proton the summation over $(i,j)$ yields a factor $-4/3$, all
three delta--functions smeared over infinitesimal regious \cite{15}
are equal to each other and scale with $\nu$ as $\nu_u^{3/2}\cdot
\delta$, where the constant $\delta$ for nucleon has been with high
accuracy computed by Green's function Monte Carlo method in \cite{15}.
As a result spin-spin interaction leads to  a contribution into
$dE_n/d\nu$ proportional to $\nu^{-3/2}$. This in turn results in the
decrease  of the quark dynamical mass $\nu_u$ by 0.035 GeV, i.e. the
contributions from the Coulomb and spin-spin interaction into the
quark mass almost exactly cancel each other. Thus we are led to a
value $\nu_u= 0.37 GeV$ given by (\ref{17a}).

\setcounter{equation}{0} \def\theequation{B.\arabic{equation}}

     \begin{center}
     {\bf Appendix B}\\

     {\bf Spin-flavour wave functions and BMM in impulse
     Approximation.
     }\\ \end{center}

     As stated in the main text we have  restricted the basis by
     considering only totally symmetric component of the baryon wave
     function in the  coordinate space (see \cite{2} for the
     discussion of corrections to this approximation). Therefore the
     spin-flavour part of the wave function has to be symmetric too.
     Then the calculation of the BMM in "impulse" (additive)
     approximation proceeds along the well trotted path \cite{2}. For
     example,  the nucleon spin-flavour
      symmetric wave function
     entering into (\ref{15}) is a combination $\psi^{symm}
     (\sigma, f)= (\varphi^{\prime\prime}(\sigma)
     \chi^{\prime\prime}(f)+ \varphi'(\sigma) \chi'(f))/\sqrt{2}$,
     where prime and double prime denote mixed symmetry functions
     symmetric and antisymmetric with respect to the $1\to 2$
     permutation. The $\Sigma^-$ spin-flavor wave function is
     obtained from that of neutron one  by substitution of the
     $u$-quark by $s$-one and so on for other baryons. Due to the
     antisymmetry of the complete wave function (\ref{15}) the
     calculation of the matrix element (\ref{14})     reduces to the
     averaging of the operator
     $(\delta_{3u}/\nu_u+\delta_{3s}/\nu_s)$ or
     $(\delta_{3d}/\nu_u+\delta_{3s}/\nu_s)$. In this way one arrives
     at the well-known relations
     $$
     \frac{\mu_n}{\mu_p}=-\frac23,
     \frac{\mu_\Lambda}{\mu_p}=-\frac{\nu_u}{3\nu_s},
     \frac{\mu_{\Sigma^+}}{\mu_p}=\frac89+\frac{\nu_u}{9\nu_s},
     \frac{\mu_{\Sigma^-}}{\mu_p}=-\frac49+\frac{\nu_u}{9\nu_s},
     $$
     \be
     \frac{\mu_{\Sigma^0}}{\mu_p}=\frac29+\frac{\nu_u}{9\nu_s},
     \frac{\mu_{\Xi^-}}{\mu_p}=\frac19-\frac49\frac{\nu_u}{\nu_s},
     \frac{\mu_{\Xi^0}}{\mu_p}=-\frac29-\frac49\frac{\nu_u}{\nu_s}
     \label{B.1}
     \ee

     Various corrections to impulse approximation have been discussed
     in the literature \cite{2},\cite{3}.

    \end{document}